# AN EXAMINATION OF THE GENERALISED POOLED BINOMIAL DISTRIBUTION AND ITS INFORMATION PROPERTIES

BEN O'NEILL[*] AND ANGUS MCLURE[**], *Australian National University*[***]

WRITTEN 8 DECEMBER 2020

**Abstract**

This paper examines the statistical properties of a distributional form that arises from pooled testing for the prevalence of a binary outcome. Our base distribution is a two-parameter distribution using a prevalence and excess intensity parameter; the latter is included to allow for a dilution or intensification effect with larger pools. We also examine a generalised form of the distribution where pools have covariate information that affects the prevalence through a linked linear form. We study the general pooled binomial distribution in its own right and as a special case of broader forms of binomial GLMs using the complementary log-log link function. We examine the information function and show the information content of individual sample items. We demonstrate that pooling reduces information content of sample units and we give simple heuristics for choosing an "optimal" pool size for testing. We derive the form of the log-likelihood function and its derivatives and give results for maximum likelihood estimation. We also discuss diagnostic testing of the positive pool probabilities, including testing for intensification/dilution in the testing mechanism. We illustrate the use of this distribution by applying it to pooled testing data on virus prevalence in a mosquito population.

POOLED TESTING; GROUP TESTING; POOLED BINOMIAL DISTRIBUTION; UNIT INFORMATION; BINOMIAL GLM; COMPLEMENTARY LOG-LOG LINK FUNCTION.

Suppose we wish to estimate the prevalence of a rare binary outcome in a large population such as infection with a rare disease, or a rare defect in a large set of items. Research problems of this kind arise in a range of contexts, including agriculture, biology, epidemiology, population health and industrial production. (Hereafter we will speak in the context of population health, where we have an "marker" affecting individuals.) Often there is a cost to administering a test for the presence of an affliction in a sample, and the rarity of the affliction means that it is prohibitively expensive to test enough individual samples to get a good inference of prevalence. An alternative method recommended by Dorfman (1943) is to group multiple individuals into larger "pools" and test these pooled samples to see if the marker is present in the larger pool (among one or more individuals). This method reduces the number of tests required for a fixed number of individuals and makes it feasible to test large samples within a limited budget. The downside is that pooling of samples leads to a loss of information — we only observe whether or not the marker is present in at least one individual in each pool. This leads to a trade-off where the researcher wishes to pool enough individuals to ensure that positive results for pools are not too rare, but also to ensure that there is not too much loss of information from pooling.

---

[*] E-mail address: ben.oneill@anu.edu.au; ben.oneill@hotmail.com.
[**] E-mail address: angus.mclure@anu.edu.au.
[***] Research School of Population Health, Australian National University, Canberra ACT 0200.

An early example of pooled testing comes from Chiang and Reeves (1962) where mosquitos are tested for the presence of a virus in order to establish an "infection rate" within a field area. The authors note that "[t]he problem of estimation is very simple indeed if one can determine for each mosquito in the sample whether or not it is infected. However, this approach to the problem is not practical, because the time and effort required to complete the isolation and identification of a virus by current laboratory techniques severely limits the number of tests that can be done. Since the number of mosquitoes that can be collected is usually large and infection rates are generally low, it is impractical to make a separate attempt at virus isolation for each individual mosquito." For small insects (e.g., mosquitos), the usual testing practice is to crush one or more insects inside a test tube and then test this for the presence or absence of the marker of interest. If the test is positive then it is known that at least one insect in the pool was infected, but when the sample contains more than insect it is impossible to determine the number and identity of the infected insects.

A related use of pooled testing is to reduce the number of tests required to identify individuals with the marker of interest. In this case, identification occurs in two or more steps where the size of pools is narrowed to narrow down or identify the positives. The simplest of these is the two-step Dorfman procedure, where the test is initially performed on a number of large pools, and then all individuals from the positive pools are retested to identify each positive individual (Dorfman 1943). The number of tests is random under the Dorfman procedure; the expected number of tests is generally less than the number of individuals. Variants of this include multi-step Dorfman procedures in which large pools that test positive may be split into a number of smaller pools for retesting, and further narrowed until a final step that tests all individuals in the positive pools. Another variant is the "array testing" procedure, in which individuals are arranged in a square or rectangular grid and pools are formed by combining samples from each row and column of the grid. Positive individuals are identified by performing individual tests on those at the intersection of positive rows and columns. If the grid sizes are appropriately designed this can further reduce the number of tests required and reduce the risk of false negatives (Phatarfod 1994; Habtesllassie 2015). These procedures are widely used in blood screening and disease identification procedures —e.g., to identify asymptomatic carriers of Covid-19 amongst blood donors (Chang 2020).

Pooled testing had been used in scientific problems prior to its introduction into the statistics literature, but it has since produced a substantial body of literature. Most examinations of the problem occur in epidemiology, population health and botany research looking at prevalence rates of rare diseases or markers in animals or plants (see e.g., Gibbs and Gower 1960; Chiang and Reeves 1962; Thompson 1962; Bhattacharyya, Karandinos and DeFoliart 1979; Walter, Hildreth and Beaty 1980; Swallow 1985). Under standard statistical assumptions, the pooled testing method leads to a sampling distribution that is a product of binomial distributions with a particular functional form for the probability of a positive pool, which can itself be regarded as a parameterised family of "pooled binomial distributions" in its own right. Theoretical treatments of this statistical problem and the resulting distributions and models is available in the literature (e.g., Chen and Swallow 1990; Hepworth 1996; Hepworth 2005). Most of this existing literature examines a simple model where the test conducted on the pools has perfect sensitivity and specificity, leading to a simple functional form for the probability of a positive pool. Existing literature shows the form of the likelihood function, score function, and Fisher information function under this standard model.

Extensions to the problem have considered the effect of imperfect sensitivity and specificity of the testing mechanism (Hwang 1976; Tu, Litvak and Pagano 1995), over-dispersion in the sampled pools (Young-Xu and Chan 2008), and the incorporation of covariate information using GLMs (Farrington 1992). In this paper we will formulate a particular example of a class of distributions arising from using a simple functional to represent dilution or intensification in the testing process for the pools. This leads us to a generalised from for the probability of a positive pool, which extends the standard model used in this field. We give a comprehensive overview of the resulting class of distribution, summarising its known properties and expanding on its information content. We show how this distribution can be used to make inferences about the level of dilution or intensification in the testing process and the prevalence of the marker of interest, and we apply this distribution to a pooled testing dataset from the literature involving pooled tests on mosquitos.

**1. Formulating the positive pool probability (PPP) function**

In prevalence testing we have a set of binary sample units that are assumed to be embedded in an exchangeable sequence. This condition allows us to invoke the "representation theorem" to establish the IID Bernoulli form for the sample units, and the prevalence parameter $0 \leq \theta \leq 1$ then corresponds to the limiting empirical frequency of the marker in the infinite sequence (see e.g., O'Neill 2009). The IID Bernoulli model occurs under an "infinite population" model, but examination of corresponding "finite population" models can be found in Bhattacharyya, Karandinos and DeFoliart (1979) and Theobald and Davie (2014). The pooled testing model occurs when instead of testing each sample unit we test pools containing multiple sample units. This manifests in a **positive pool probability (PPP) function** that determines the probability of a positive test for a pool based on the underlying **prevalence parameter** $\theta$, the **pool size** $s$, and any other parameters included in the analysis. Throughout this paper we will denote values of this function for the base parameters as $\phi_s(\theta)$, but we will add other model parameters to the function as required, or re-parameterise to transformed parameters as required.

The simplest form of the positive pool probability function arises when we assume that the test has perfect sensitivity and specificity for pools of any size —i.e., those pools containing no individuals with the marker of interest always return a negative test result and those pools with one or more individuals with the marker always return a positive test result. In this case the positive pool probability function is simply the probability that a binomial random variable of size $s$ and probability $\theta$ is non-zero:

$$\phi_s(\theta) = \sum_{\ell=1}^{s} \binom{s}{\ell} \theta^\ell (1-\theta)^{s-\ell} = 1 - (1-\theta)^s.$$

There are two natural ways to generalise this model to allow for imperfect tests. One method is to assume that the probability of a positive test in a pool is fully determined by the size of the pool and the number of positive individuals in that pool (i.e., the result of the test in a pool is conditionally independent of the prevalence given these values). To do this, we let $h(s, \ell)$ be the probability of a positive result in a pool of size $s$ containing $0 \leq \ell \leq s$ positive cases. Under this formulation the positive pool probability function is:

$$\phi_s(\theta) = \sum_{\ell=0}^{s} \binom{s}{n} \theta^\ell (1-\theta)^{s-\ell} \cdot h(s, \ell).$$

For any pool size $s \in \mathbb{N}$, the requirement of perfect specificity corresponds to the condition $h(s, 0) = 0$ and the requirement of perfect sensitivity corresponds to the condition $h(s, \ell) = 1$ for all $\ell > 0$. In order for the probability function $h$ to be sensible, we impose monotonicity requirements on the function, requiring that it be non-decreasing in $\ell$ and non-increasing in $s$. If sensitivity decreases with $s$ we have a "dilution" effect and if specificity decreases with $s$ we have an "intensification" effect.

The second natural way to generalise the positive pool probability function for imperfect tests is to allow the "effective pool size" to be different to the actual pool size. In the general case, we can let $g(s)$ be the effective pool size for a pool with actual size $s$. Then the positive pool probability function is formulated as:

$$\phi_s(\theta) = 1 - (1-\theta)^{g(s)}.$$

If the weighting ratio $g(s)/s$ is a decreasing function of $s$ then larger pool sizes "dilute" the efficacy of the test. On the other hand, if the weighting ratio $g(s)/s$ is an increasing function of $s$ then larger pool sizes "intensify" the test. A simple choice of the effective pool size $g$ that allows both possibilities with a single parameter is the function $g(s) = s^{1+\lambda}$, which leads to a positive pool probability function:[1]

$$\phi_s(\theta, \lambda) = 1 - (1-\theta)^{s^{1+\lambda}}.$$

In this formulation, the parameter $\lambda \geq -1$ gives a generalisation from the model for a perfect test. In the special case where $\lambda = 0$ we recover the model for a perfect test. If $-1 \leq \lambda < 0$ we model a test with "dilution" and if $\lambda > 0$ we model a test with "intensification". For this reason, we will call $\lambda$ the **excess intensity parameter**.

These two approaches to formulating the positive pool probability function give rise to general functional forms that can incorporate various kinds of departures from perfect testing. It is worth noting that not every functional form for the positive pool probability is compatible with both of these methods, and it requires some judgment to determine how best to formulate an appropriate function. The class of models defined by the set of possible choices of $h$ (using the first method shown above) includes all models where the probability of a positive test depends only on the size of the pool and the number of positive individuals. Moreover, since the two methods we have considered do not lead to equivalent models except in trivial cases, models of the latter type implicitly assume that the test characteristic depends on the prevalence $\theta$ beyond its effect on the number of positive individuals in the pool (i.e., forms from this method do not always satisfy the conditional independence assumption of the first method). Violation of this conditional independence requirement may seem like a significant drawback, but that condition is not always realistic. For instance, in macroparasitic infections, the mean burden of parasites amongst infected individuals is often higher in populations where the prevalence of infection is higher. Consequently, pooled tests with a given size and containing a given number of infected individuals may contain a higher concentration of the marker of interest in populations with higher prevalence of infection, leading to a higher probability of a positive test. The nature of this causal and statistical dependence may be very complex, and it is difficult to test these effects separately from the analysis of prevalence. Consequently, the second method shown above gives a way to simplify this complexity by directly stipulating the "effective" pool size that accrues to any actual pool size. The ability to give a simple model of "dilution" and "intensification" is particularly convenient. (Note also that there are some other *ad hoc* methods that have been used to obtain the positive pool probability function; see e.g., Hwang 1976.)

---

[1] There is one sense in which this choice of $g$ is the "simplest" choice of function here. If we look at the pool size and effective pool size on the log scale (to allow an unconstrained analysis) we get $\log g(s) = (1 + \lambda) \log s$ so that our function gives a linear form for the effective pool size function. As is known from the theory of Taylor series, the linear function is the first-order approximation to a broad class of non-periodic functions, and so we can regard this choice of $g$ to be the natural "first-order" form for the effective pool size.

In the remainder of the paper we will formulate a class of pooled binomial distributions using the positive pool probability function $\phi_s(\theta, \lambda) = 1 - (1-\theta)^{s^{1+\lambda}}$ using the excess intensity parameter $\lambda \geq -1$. This form provides a single parameter model of dilution and intensification that is particularly convenient for regression analyses and allows simple hypothesis testing for the presence of these departures from perfect testing.

## 2. The pooled binomial distribution and its representation

The pooled testing model can be described using a vector $\mathbf{z} = (z_1, \ldots, z_K)$ representing the test outcomes for $K$ pools of sizes $\tilde{\mathbf{s}} = (\tilde{s}_1, \ldots, \tilde{s}_K)$. This leads to the sampling distribution:

$$p(\mathbf{z}|\tilde{\mathbf{s}}, \theta) = \prod_{k=1}^{K} \text{Bern}(z_k | \phi_{\tilde{s}_k}(\theta)).$$

Since there may be multiple pools with the same pool size, an alternative representation uses the size vector $\mathbf{n} = (n_1, \ldots, n_M)$ and pool size vector $\mathbf{s} = (s_1, \ldots, s_M)$ to encapsulate the pools, and the count vector $\mathbf{y} = (y_1, \ldots, y_M)$ to count the number of positive pools. These vectors are related to the underlying test outcomes by the equations:

$$\sum_{i=1}^{M} n_i \cdot \mathbb{I}(s_i = s) = \sum_{k=1}^{K} \mathbb{I}(\tilde{s}_k = s),$$

$$\sum_{i=1}^{M} y_i \cdot \mathbb{I}(s_i = s) = \sum_{k=1}^{K} z_k \cdot \mathbb{I}(\tilde{s}_k = s).$$

Since the underlying $z_k$ values are independent Bernoulli random variables, the elements of **y** are independent random variables with $y_i \sim \text{Bin}(n_i, \phi_{s_i}(\theta, \lambda))$, which gives us the distribution $\mathbf{y} \sim \text{PoolBin}(\mathbf{n}, \mathbf{s}, \theta, \lambda)$ defined below.

**DEFINITION (Pooled binomial distribution):** This distribution has the mass function:

$$\text{PoolBin}(\mathbf{y}|\mathbf{n}, \mathbf{s}, \theta, \lambda) = \prod_i \text{Bin}(y_i|n_i, 1-(1-\theta)^{s_i^{1+\lambda}})$$

$$= (1-\theta)^{\sum_i (n_i - y_i) s_i^{1+\lambda}} \prod_i \binom{n_i}{y_i} (1-(1-\theta)^{s_i^{1+\lambda}})^{y_i},$$

where **n** is the **pool-count vector**, **s** is the **pool-size vector**, $0 \leq \theta \leq 1$ is the **prevalence** and $\lambda \geq -1$ is the **excess intensity**. This is a generalisation of the binomial distribution.[2]

**REMARK:** It is possible to parameterise this distribution without an explicit pool-size vector **s**. To do this we can use implicit pool-sizes $\mathbf{s} = (1, 2, 3, \ldots, \bar{s})$ up to a maximum pool-size $\bar{s}$ and use vectors $\mathbf{y} = (y_1, \ldots, y_{\bar{s}})$ and $\mathbf{n} = (n_1, \ldots, n_{\bar{s}})$ where the index refers directly to the pool-size. One can also use infinite sequences **y** and **n** with counts running across all positive integer pool sizes; this ensures that there is no need to change the representation when the sample size changes, including when conducting asymptotic analysis. □

In many applications of interest, **n** and **s** are fixed by the sampling design. Under this condition the vector **y** is a sufficient statistic for the remaining parameters in our sampling method. To see this, we note that the Fisher-Neyman factorisation of the mass of **z** is:

$$p(\mathbf{z}|\mathbf{s}, \theta) = \prod_{k=1}^K \text{Bern}(z_k|\phi_{s_k}(\theta)) = \prod_{k=1}^K \phi_{s_k}(\theta, \lambda)^{z_k}(1-\phi_{s_k}(\theta, \lambda))^{1-z_k}$$

$$= \prod_i \phi_{s_i}(\theta, \lambda)^{y_i}(1-\phi_{s_i}(\theta, \lambda))^{n_i - y_i}$$

$$= \prod_i (1-(1-\theta)^{s_i^{1+\lambda}})^{y_i}((1-\theta)^{s_i^{1+\lambda}})^{n_i - y_i}$$

$$= (1-\theta)^{\sum_i (n_i - y_i) s_i^{1+\lambda}} \prod_i (1-(1-\theta)^{s_i^{1+\lambda}})^{y_i}$$

$$\overset{\theta,\lambda}{\propto} \text{PoolBin}(\mathbf{y}|\mathbf{n}, \mathbf{s}, \theta, \lambda).$$

---

[2] In the case where $\mathbf{n} = n$ and $\mathbf{s} = 1$ we have $\text{PoolBin}(\mathbf{n}, \mathbf{s}, \theta, \lambda) = \text{Bin}(n, \theta)$ so the pooled binomial distribution generalises the binomial distribution.

Minimal sufficiency occurs when we reduce the vectors so that **s** has distinct elements. (If **s** has any repeat elements then we can aggregate the relevant pool counts in **n** and positive pool counts in **y** without loss of information.) In the special case where $\lambda = 0$ the probability mass function for the pooled binomial distribution is a polynomial of degree $N$ in $\theta$.[3]

For estimation purposes it is often useful to parameterise the pooled binomial distribution in terms of a transformed parameter on the extended real numbers. Here we transform using the **complementary log-log function** $\text{CLL}(\theta) = \log(-\log(1-\theta))$ that maps probability values $0 \leq \theta \leq 1$ to the extended real numbers.[4] This function is strictly increasing and its inverse is the **inverse complementary log-log function** $\text{ICLL}(\eta) = 1 - \exp(-\exp(\eta))$ that maps the extended real numbers back to probability values. The benefit of the complementary log-log function in the present context is that it simplifies the positive pool probabilities as follows:

$$\text{CLL}(\phi_s(\theta, \lambda)) = (1+\lambda)\log(s) + \text{CLL}(\theta).$$

This form means that the complementary log-log function can be applied to the positive pool probabilities to separate the pool size from the prevalence parameter. We also see in this form that the total intensity (one plus excess intensity) forms the "slope coefficient" for the effect of the log-pool-size. (Note also that the other formulations of the dilution effect considered earlier do not have this property.) Using the transformed parameter $\eta = \text{CLL}(\theta)$ the probability mass function for the pooled binomial distribution can be written more succinctly as:[5]

$$\text{PoolBin}(\mathbf{y}|\mathbf{n}, \mathbf{s}, \eta, \lambda) = \exp\left(-\exp(\eta) \sum_i (n_i - y_i) s_i^{1+\lambda}\right)$$

$$\times \prod_i \binom{n_i}{y_i} (1 - \exp(-\exp(\eta) s_i^{1+\lambda}))^{y_i}.$$

The transformed parameter is useful for a number of purposes, including numerically stable computation of the MLE, and modelling of pooled testing data using covariates. The present parameterisation of the mass function uses the CLL function as a "link" function that operates on the positive pool probability to separate the pool size from the prevalence parameter. Later we will examine broader models involving covariates, but we can already see that the pool size $s$ is separated from the prevalence parameter under this "link" function and so the pool size is essentially treated as a covariate that affects the positive pool probability.

---

[3] In this case it is possible to use the binomial theorem and other related mathematical results to expand the polynomial and compute its coefficients. The polynomial can then be written in expanded form (i.e., as a sum of powers of $\theta$) but this form is cumbersome and unilluminating.
[4] We take $\text{CLL}(0) = -\infty$ and $\text{CLL}(1) = +\infty$ so that the domain of the function includes all probability values.
[5] To avoid confusion, note that we use a slight abuse of notation here by using generic function notation —i.e., the function $\text{PoolBin}(\mathbf{y}|\mathbf{n}, \eta)$ has a different form than the function $\text{PoolBin}(\mathbf{y}|\mathbf{n}, \theta)$.

We need not bother to derive the moments of the pooled binomial distribution, since they are trivially related to the moments of the underlying binomial distribution. The elements of **y** in the pooled binomial distribution are independent, so the moment generating function for the pooled binomial distribution is a product of the moment generating functions for the binomial distribution, using the probabilities of positive pools of the relevant sizes. (Other generating functions are also trivially related to the generating functions for the binomial distribution.) The mean vector and variance matrix for the observable vector **y** is easily obtained from the fact that its elements are independent binomial random variables.

## 3. Information properties of the pooled binomial distribution

Pooled testing is motivated by a desire to reduce testing costs, allowing for sampling of more units in the analysis of binary data. In view of this objective, the information properties of the distribution are of particular interest. One expects pooled testing to reduce the information in the sample compared to unit testing, but this may be justified as a trade-off against the expense of testing individual sample units. In order to formally assess this aspect of the distribution we will examine the Fisher information for the distribution. The information matrix is also useful because it determines the asymptotic variance of the MLE of the parameters, and so we will also use it later when we look at parameter estimation.

We defer the derivation of the Fisher information matrix until a later section (where we look at the log-likelihood function for the distribution). The matrix is given by:[6]

$$\mathbf{I}(\theta, \lambda) = \begin{bmatrix} \dfrac{1}{(1-\theta)^2} \sum_i r_i(\theta, \lambda) & \log^2(1-\theta) \sum_i r_i(\theta, \lambda) \log^2(s_i) \\ \log^2(1-\theta) \sum_i r_i(\theta, \lambda) \log^2(s_i) & -\dfrac{\log(1-\theta)}{1-\theta} \sum_i r_i(\theta, \lambda) \log(s_i) \end{bmatrix},$$

where the terms $r_i(\theta, \lambda)$ used in the summation are given by:

$$r_i(\theta, \lambda) \equiv n_i s_i^{2(1+\lambda)} \cdot \dfrac{1 - \phi_{s_i}(\theta, \lambda)}{\phi_{s_i}(\theta, \lambda)}.$$

---

[6] Here $\log^2(s) \equiv (\log(s))^2$ (i.e., the log-squared function), **not** $\log\log(s)$ (the log-log function).

In practical applications of pooled testing, the prevalence parameter $\theta$ is the object of interest in our inferences and the excess intensity $\lambda$ is merely a nuisance parameter. We will therefore focus on the information with respect to the parameter $\theta$ which is given by the top-left entry of the Fisher information matrix. If we note that there are $n_i s_i$ sample units in the $i$th element of the data vectors, we can rewrite this measure of information as:

$$\mathbf{I}_{\theta\theta}(\theta, \lambda) = \sum_i n_i s_i \cdot I_{s_i}(\theta, \lambda) = \sum_{k=1}^{K} s_k \cdot I_{s_k}(\theta, \lambda),$$

where the information contribution from each individual sample unit present in a pool of size $s$ is given by the **unit information** function:

$$I_s(\theta, \lambda) = \frac{s^{2\lambda+1}(1-\theta)^{s^{1+\lambda}-2}}{1-(1-\theta)^{s^{1+\lambda}}}.$$

This function measures the information content (pertaining to the prevalence parameter $\theta$) of a single sample unit based on the pool-size of the pool it is placed in. This allows us to look at the information loss that occurs from pooled testing.

We begin with the simple case where the pooled testing mechanism has no intensification or dilution (i.e., when $\lambda = 0$). In this case it is possible to show that the unit information is strictly decreasing in $s$ for all $\theta > 0$ (and is constant when $\theta = 0$) (see Appendix). This means that units placed in pools of larger size give less information about the prevalence parameter $\theta$ than units placed in pools of smaller size. This accords with our intuition, since pooled testing entails a loss of information about the outcomes of the underlying binary sample units, and the larger the pool the larger the loss of information.

Figure 1 below shows the unit information function for pool sizes up to five, in the case where there is no intensification or dilution. Unit testing (pool size $s = 1$) reduces to the standard information function for the Bernoulli distribution (which is symmetric in $\theta$). Unit information decreases as $s$ increases, but is reduced most heavily for larger values of $\theta$ (Figure 1). This is in accordance with our intuition that pooled testing is most useful when examining a rare marker, since it is unlikely that more than one sample unit in a small pool will be positive.

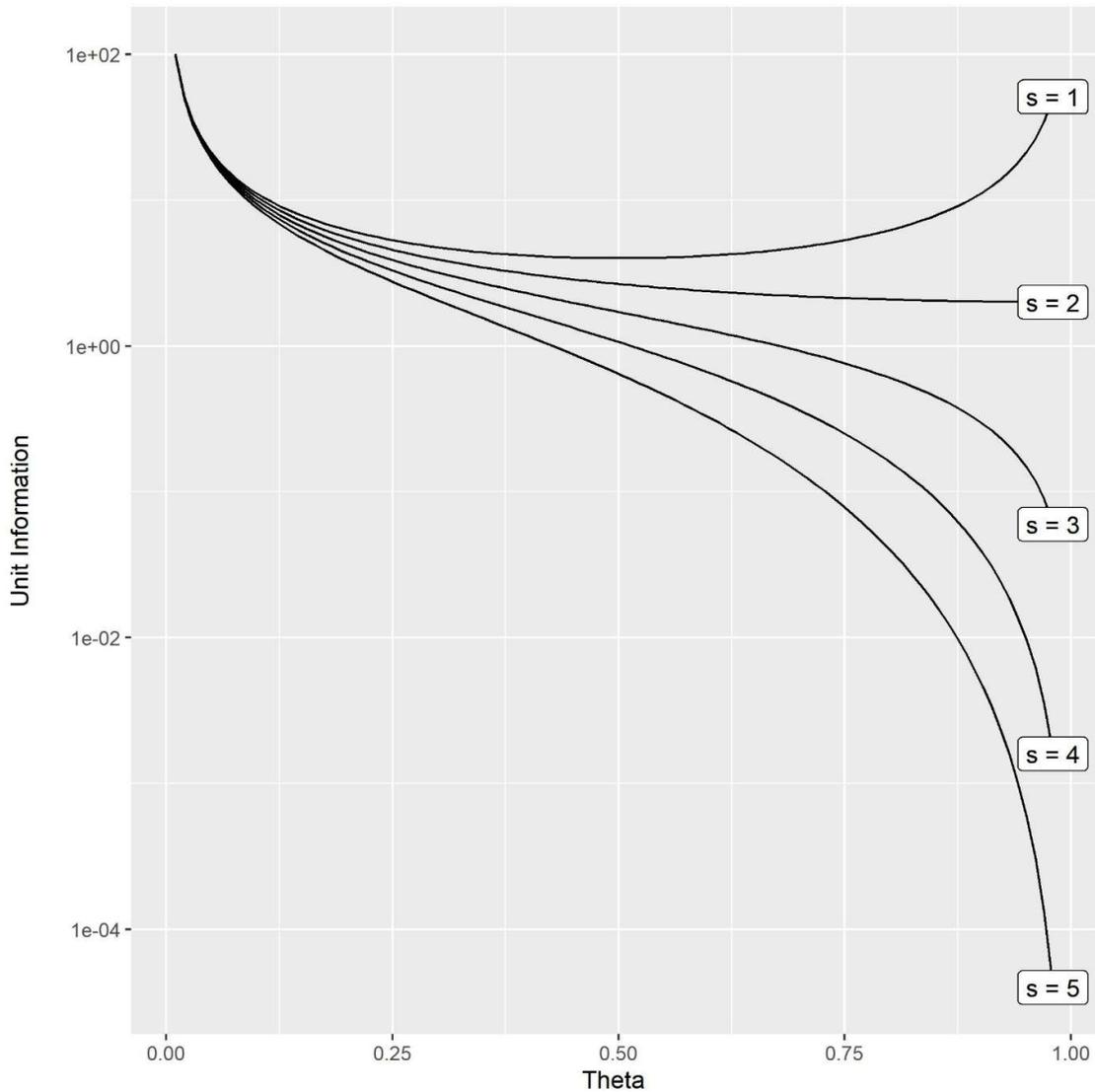

**FIGURE 1:** Unit information function for a unit in a pool of size $s$ (when $\lambda = 0$)

Since any pooling represents some loss of information, it is only worth considering if there is some countervailing impediment or cost for testing each unit. In such cases there is a trade-off between the testing cost and the loss of information from pooling. One simple way to the analyse this trade-off is to consider the information-per-unit-cost for pools of different sizes. Suppose we face two costs for our experimental design: a sampling cost $a \geq 0$ per unit (which is sometimes a sunk cost) and a testing cost $t \geq 0$ per test. This means that a single sample unit in a pool of size $s$ costs $a + t/s$ and gives us unit information $I_s(\theta, \lambda)$, so the information-per-unit-cost in a pool of size $s$ is:

$$\text{IPUC}_s(\theta, \lambda) = \frac{I_s(\theta, \lambda)}{a + t/s} = \frac{1}{as + t} \cdot \frac{s^{2(1+\lambda)}(1-\theta)^{s^{1+\lambda}-2}}{1 - (1-\theta)^{s^{1+\lambda}}}.$$

For any parameter values the value $\hat{s}(\theta, \lambda) = \text{argmax}_s \text{ IPUC}_s(\theta, \lambda)$ is the pool size giving the highest information-per-unit-cost (the "optimal" pool size). In the simple case where there is no intensification or dilution of the testing mechanism (i.e., when $\lambda = 0$) it can be shown that this function $\hat{s}(\theta)$ is decreasing in $\theta$, so a smaller prevalence gives a higher optimal pool size. Figure 2 shows the optimal pool size when there is no intensification or dilution of the testing mechanism and the sampling cost is zero (e.g., when the sampling cost is a "sunk cost").

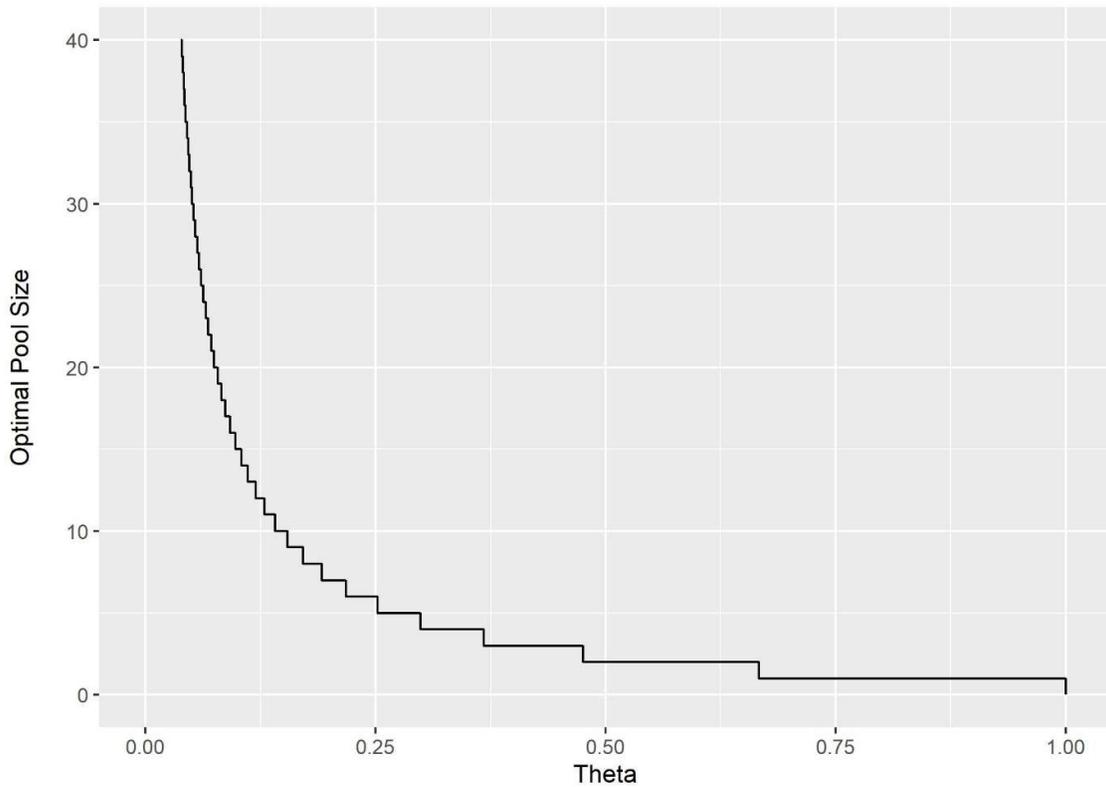

**FIGURE 2:** Pool size maximising information-per-unit-cost (when $\lambda = 0$, and $a = 0$)

Since the information function determines the asymptotic variance of the MLE it is also useful to examine the information for the alternative parameter $\eta = \text{CLL}(\theta)$. For any differentiable injective transformation $\mathbf{g}$ we have information $\mathbf{I}(\mathbf{g}(\theta, \lambda)) = \mathbf{J}(\theta, \lambda)^\text{T} \mathbf{I}(\theta, \lambda) \mathbf{J}(\theta, \lambda)$ where we use the Jacobian matrix $\mathbf{J}(\theta, \lambda) = d(\theta, \lambda)/d\mathbf{g}(\theta, \lambda)$ (see, e.g., Cox and Hinkley 1974, p. 109). We have $\text{CLL}'(\theta) = 1/(1-\theta) \log(1-\theta)$ and $\log(1-\theta) = -\exp(\eta)$ which gives:

$$\mathbf{I}(\eta,\lambda) = \begin{bmatrix} \exp(2\eta)\sum_i r_i(\theta,\lambda) & -\exp(3\eta - \exp(\eta))\sum_i r_i(\theta,\lambda)\log^2(s_i) \\ -\exp(3\eta - \exp(\eta))\sum_i r_i(\theta,\lambda)\log^2(s_i) & \exp(\eta + \exp(\eta))\sum_i r_i(\theta,\lambda)\log(s_i) \end{bmatrix},$$

where the terms $r_i(\eta,\lambda)$ used in the summation are given by:

$$r_i(\eta,\lambda) \equiv n_i s_i^{2(1+\lambda)} \cdot \frac{\exp(-s_i^{1+\lambda}\exp(\eta))}{1 - \exp(-s_i^{1+\lambda}\exp(\eta))}.$$

This gives the corresponding unit information:

$$\text{IPUC}_s(\eta,\lambda) = \exp(3\eta) \cdot \frac{s^{2\lambda+1}\exp(-s^{1+\lambda})}{1 - \exp(-s^{1+\lambda}\exp(\eta))}.$$

Cases where there is dilution or intensification in the testing mechanism are shown in Figure 3 below, where we show plots over several different values for the intensity parameter. As can be seen from the figure, for each fixed pool size larger than one, higher intensity means that the unit information decreases more rapidly with respect to the prevalence parameter. When there is high excess intensity and the pool size is not small, the unit information decreases rapidly as the prevalence increases. Intuitively, this reflects the fact that intensification in the testing mechanism means that the "effective pool size" is much higher than the actual pool size, and so the rapid decrease in the unit information with respect to the prevalence parameter is just as would be expected with a higher pool size and no intensification.

In cases where there is dilution or intensification of the testing mechanism the change in the "optimal" pool size is complicated and non-monotonic. Whilst higher intensity leads to a more rapid decrease in the unit information with respect to the prevalence, it is also the case that for a fixed prevalence value, the change in the unit information at a given pool size may occur in either direction. From Figure 3 we can see that for higher intensity values the unit information lines for different pool sizes cross at relatively low values of the prevalence parameter, which reflects the fact that the unit information can now be higher with higher pool sizes. This affects the "optimal" pool size in a way that depends on the value of the prevalence parameter; for large values of the prevalence parameter the unit information will be extremely small for high pool sizes and so the optima will tend to shift lower (though in discrete increments); for small values of the prevalence parameter the unit information will be larger for high pool sizes and so the optimal will tend to shift higher (though again in discrete increments).

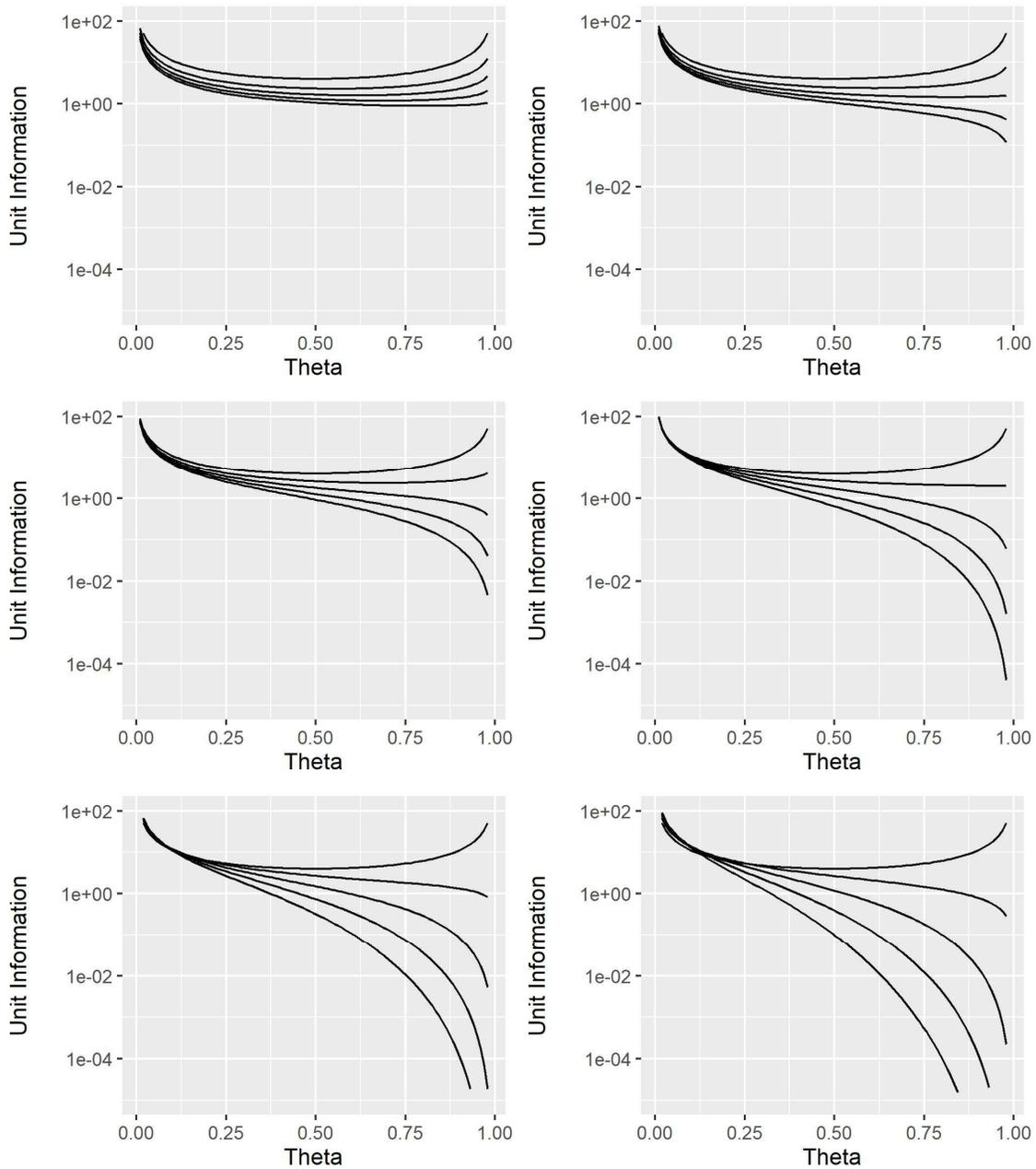

**FIGURE 3:** Unit information function for a unit in a pool of sizes $s = 1, 2, 3, 4, 5$
(top-left to bottom-right show the cases $\lambda = -0.6, -0.4, -0.2, 0.0, 0.2, 0.4$)

To deal with the possibility of model error, it is important to conduct diagnostic analysis by comparing empirical estimates of the positive pool probabilities to the theoretical "curve" of the function $\phi_s(\theta, \lambda)$ over the pool sizes $s$. Substantial departure of the empirical estimates for individual pools from the curve constitutes evidence that the stipulated form for the function is inadequate, due to failures of the underlying model assumptions. Thus, while there may be a single "optimal" pool size in terms of the information-per-unit-cost for $\theta$ we stress that it is generally bad practice to conduct pooled testing with only a single pool size, or only small variation in pool size, since this precludes effective diagnostic testing. Moreover, if the likely position of the prevalence parameter is highly variable *a priori* then using a range of pool sizes that are "optimal" against different prevalence values will tend to give at least some pools that give high levels of information about the prevalence parameter (Gu et al. 2004). More detailed analysis of pool design can be found in Remlinger et al (2006) and Schaarschmidt (2007).

Good statistical practice for pooled testing should involve some reasonable amount of variation in the pool sizes. The "optimal" pool size should be used as a rule-of-thumb for the mean pool size, with variation intentionally incorporated into the experimental design. The appropriate amount of variation in the pool sizes depends on the trade-off between information optimisation versus the capacity to diagnose model error. An ancillary benefit is that variation in the pool size gives robustness against sub-optimal pool size choice due to poor prior estimation of the prevalence parameter. We examine diagnostic testing for the distribution later in this paper, but for now it suffices to say that this analysis can only be done if there is enough data over multiple different pools to diagnose departures from the theoretical model.

**4. Likelihood analysis and the maximum-likelihood estimator**

It is useful to examine the log-likelihood function and maximum-likelihood estimator (MLE) for the prevalence parameter using data from the pooled binomial distribution. It will be shown that the log-likelihood function is strictly quasi-concave (therefore so is the likelihood function) and so the MLE occurs at the unique critical point of the log-likelihood function. This leads to an implicit critical point equation for the MLE that can be computed using Newton iteration. In this section we show the relevant properties of the log-likelihood function and recommend a method for efficient and stable optimisation.

For simplicity, we begin by looking at the simplest case where we have a single observed vector **y** from the pooled binomial distribution. In this case, the log-likelihood function for the pooled binomial distribution is given by:

$$\ell_{\mathbf{y}}(\theta, \lambda) = \text{const} + \log(1-\theta) \sum_i (n_i - y_i) s_i^{1+\lambda} + \sum_i y_i \log \phi_{s_i}(\theta, \lambda).$$

The first-order partial derivatives of the log-likelihood function are:

$$\frac{\partial \ell_\mathbf{y}}{\partial \theta}(\theta, \lambda) = -\frac{1}{1-\theta} \sum_i s_i^{1+\lambda} \left( n_i - \frac{y_i}{\phi_{s_i}(\theta, \lambda)} \right),$$

$$\frac{\partial \ell_\mathbf{y}}{\partial \lambda}(\theta, \lambda) = \log(1-\theta) \sum_i s_i^{1+\lambda} \log(s_i) \left( n_i - \frac{y_i}{\phi_{s_i}(\theta, \lambda)} \right).$$

The second-order partial derivatives of the log-likelihood function are:

$$\frac{\partial^2 \ell_\mathbf{y}}{\partial \theta^2}(\theta, \lambda) = -\frac{1}{(1-\theta)^2} \sum_i s_i^{1+\lambda} \left( n_i - \frac{y_i}{\phi_{s_i}(\theta, \lambda)} \left[ 1 - \frac{s_i^{1+\lambda}(1-\theta)^{s_i^{1+\lambda}}}{\phi_{s_i}(\theta, \lambda)} \right] \right),$$

$$\frac{\partial^2 \ell_\mathbf{y}}{\partial \theta \partial \lambda}(\theta, \lambda) = -\frac{1}{1-\theta} \sum_i s_i^{1+\lambda} \log(s_i) \left( n_i - \frac{y_i}{\phi_{s_i}(\theta, \lambda)} \left[ 1 + \frac{s_i^{1+\lambda} \log(1-\theta)(1-\theta)^{s_i^{1+\lambda}}}{\phi_{s_i}(\theta, \lambda)} \right] \right),$$

$$\frac{\partial^2 \ell_\mathbf{y}}{\partial \lambda^2}(\theta, \lambda) = \log(1-\theta) \sum_i s_i^{1+\lambda} \log^2(s_i) \left( n_i - \frac{y_i}{\phi_{s_i}(\theta, \lambda)} \left[ 1 + \frac{s_i^{1+\lambda} \log(1-\theta)(1-\theta)^{s_i^{1+\lambda}}}{\phi_{s_i}(\theta, \lambda)} \right] \right).$$

The second derivatives can be written using the first derivatives as:

$$\frac{\partial^2 \ell_\mathbf{y}}{\partial \theta^2}(\theta, \lambda) = \frac{1}{1-\theta} \left[ \frac{\partial \ell_\mathbf{y}}{\partial \theta}(\theta) - \frac{1}{1-\theta} \sum_i s_i^{2(1+\lambda)} y_i \cdot \frac{1 - \phi_{s_i}(\theta, \lambda)}{\phi_{s_i}(\theta, \lambda)^2} \right],$$

$$\frac{\partial^2 \ell_\mathbf{y}}{\partial \theta \partial \lambda}(\theta, \lambda) = \frac{1}{1-\theta} \left[ -\frac{1}{\log(1-\theta)} \frac{\partial \ell_\mathbf{y}}{\partial \lambda}(\theta, \lambda) + \log(1-\theta) \sum_i s_i^{2(1+\lambda)} y_i \log(s_i) \cdot \frac{1 - \phi_{s_i}(\theta, \lambda)}{\phi_{s_i}(\theta, \lambda)^2} \right],$$

$$\frac{\partial^2 \ell_\mathbf{y}}{\partial \lambda^2}(\theta, \lambda) = \frac{\partial \ell_\mathbf{y}}{\partial \lambda}(\theta, \lambda) + \log(1-\theta) \sum_i s_i^{1+\lambda} \log(s_i) [\log(s_i) - 1] \left( n_i - \frac{y_i}{\phi_{s_i}(\theta, \lambda)} \right)$$

$$- \log^2(1-\theta) \sum_i s_i^{2(1+\lambda)} y_i \log^2(s_i) \cdot \frac{1 - \phi_{s_i}(\theta, \lambda)}{\phi_{s_i}(\theta, \lambda)^2}.$$

The above equations allow us to determine the critical point equation for joint estimation of the two parameters or conditional estimation of the prevalence. When estimating both the prevalence parameter and excess intensity parameter jointly, we solve the joint critical point equation:

$$\sum_i s_i^{1+\hat{\lambda}} \left( n_i - \frac{y_i}{\phi_{s_i}(\hat{\theta}, \hat{\lambda})} \right) = \sum_i s_i^{1+\hat{\lambda}} \log(s_i) \left( n_i - \frac{y_i}{\phi_{s_i}(\hat{\theta}, \hat{\lambda})} \right) = 0.$$

When estimating the prevalence for a fixed excess intensity parameter $\lambda$, we solve the equation:

$$\sum_i s_i^{1+\lambda} \left( n_i - \frac{y_i}{\phi_{s_i}(\hat{\theta}, \lambda)} \right) = 0.$$

At any critical point value $\hat{\theta}$ we have $\partial \ell_{\mathbf{y}}/\partial \theta \, (\hat{\theta}, \lambda) = 0$ and $\partial^2 \ell_{\mathbf{y}}/\partial \theta^2 \, (\hat{\theta}, \lambda) < 0$. Using a simple variation of the "only critical point in town" argument, it follows that for any fixed $\lambda$ the log-likelihood is a strictly quasi-concave function in $\theta$ with a unique critical point that is its global maximising point. The critical point equation does not have a closed form solution (except in some trivial special cases) so it is maximised numerically. In the trivial case where there is only one pool size $s$, we can see that the MLE for the prevalence parameter reduces to the simple closed-form estimator $\hat{\theta} = 1 - (1 - y/n)^{s^{-(1+\lambda)}}$.

For some purposes it is useful to reframe the equations for the pooled binomial distribution in terms of simple estimators relating to the underlying counts. Let $n \equiv \sum_s n_s$ denote the total number of pools in the data and define the functions $\hat{p}_n(s) = n_s/n$ and $\hat{\phi}_{n,s} = y_s/n_s$. The function $\hat{p}_n$ encapsulates the sampling design by showing the distribution of the pool sizes over the pools in the analysis.[7] The estimator $\hat{\phi}_{n,s}$ is the empirical estimator for the probability of a positive result in pool $s$, using only the data from that pool. The latter is the MLE for the positive pool probability if we ignore the information in the other pools. Using these functions, we can rewrite the derivatives of the log-likelihood function in a more compact and intuitive form. In particular, the critical point equation for the MLE of $\theta$ can be written as:

$$\sum_i s_i^{1+\lambda} \hat{p}_n(s) \left( 1 - \frac{\hat{\phi}_{n,s}}{\phi_s(\hat{\theta}, \lambda)} \right) = 0.$$

The constrained functional form of $\phi_s$ means that we cannot generally equate the empirical estimators with the estimated positive pool probabilities, but we can see that the MLE involves an attempt to ensure that the empirical estimators $\hat{\phi}_{n,s}$ are close to the estimated positive pool probabilities, with the closeness being weighed over the pools in the data.

For numerical stability of the optimisation (and for simpler interaction with covariate models) we recommend maximising the log-likelihood with respect to the unconstrained real parameter $\eta = \text{CLL}(\theta)$. With this transformation the log-likelihood function is given by:[8]

$$\ell_{\mathbf{y}}(\eta, \lambda) = \text{const} - \exp(\eta) \sum_i (n_i - y_i) s_i^{1+\lambda} + \sum_i y_i \log1\text{mexp}(s_i^{1+\lambda} \exp(\eta)).$$

---

[7] In most contexts this is not actually an estimator since the values $n_s$ are fixed by the sample design. If the pool sizes are IID random variables then this is the empirical estimator of their distribution.
[8] Here we use the function $\log1\text{mexp}(x) = \log(1 - \exp(-x))$. This function is programmed directly into most mathematical and statistical software.

The first partial derivatives of the log-likelihood function with respect to this parameter are:[9]

$$\frac{\partial \ell_{\mathbf{y}}}{\partial \eta}(\eta, \lambda) = -\exp(\eta) \sum_i s_i^{1+\lambda}\left(n_i - \frac{y_i}{\phi_{s_i}(\eta, \lambda)}\right),$$

$$\frac{\partial \ell_{\mathbf{y}}}{\partial \lambda}(\eta, \lambda) = -\exp(\eta) \sum_i s_i^{1+\lambda}\log(s_i)\left(n_i - \frac{y_i}{\phi_{s_i}(\eta, \lambda)}\right),$$

and the second partial derivatives are:

$$\frac{\partial^2 \ell_{\mathbf{y}}}{\partial \eta^2}(\eta, \lambda) = -\exp(\eta) \sum_i s_i^{1+\lambda}\left(n_i - \frac{y_i}{\phi_{s_i}(\eta, \lambda)}\left[1 - s_i^{1+\lambda}\exp(\eta)\cdot\frac{1-\phi_{s_i}(\eta, \lambda)}{\phi_{s_i}(\eta, \lambda)}\right]\right),$$

$$\frac{\partial^2 \ell_{\mathbf{y}}}{\partial \eta \partial \lambda}(\eta, \lambda) = -\exp(\eta) \sum_i s_i^{1+\lambda}\log^2(s_i)\left(n_i - \frac{y_i}{\phi_{s_i}(\eta, \lambda)}\left[1 - \exp(\eta)\cdot\frac{1-\phi_{s_i}(\eta, \lambda)}{\phi_{s_i}(\eta, \lambda)}\right]\right),$$

$$\frac{\partial^2 \ell_{\mathbf{y}}}{\partial \lambda^2}(\eta, \lambda) = -\exp(\eta) \sum_i s_i^{1+\lambda}\log^2(s_i)\left(n_i - \frac{y_i}{\phi_{s_i}(\eta, \lambda)}\left[1 - s_i^{1+\lambda}\exp(\eta)\cdot\frac{1-\phi_{s_i}(\eta, \lambda)}{\phi_{s_i}(\eta, \lambda)}\right]\right).$$

Since reparameterization to $\eta$ is done using a strictly increasing function, strict quasi-concavity of the log-likelihood function with respect to $\eta$ is preserved, so that for any value of $\lambda$ the log-likelihood has a unique critical point $\hat{\eta}$ that is its global maximising point.

Under this reparameterisation it is important to note that the MLE $\hat{\eta}$ may be at infinity or minus infinity, corresponding to prevalence MLEs of zero or one (and these MLEs correspond to data values $\mathbf{y} = \mathbf{0}$ and $\mathbf{y} = \mathbf{n}$ respectively). In practice it is easy to deal with this by checking for these data values as special cases. In other cases, optimising over real $\eta$ using Newton iteration is stable and efficient. We recommend starting the iterations at a crude closed form estimator suggested by Le (1980), which assumes that all pools have the mean pool size $\bar{s}_n = N/n$. Taking the starting excess intensity $\hat{\lambda} = 0$ and solving $N_+/N = \phi_{\bar{s}_n}(\hat{\theta}, 0)$ leads to the estimator $\hat{\theta} = 1 - (1 - N_+/N)^{n/N}$ which then gives the starting value $\hat{\eta} = \log(-n\log(1 - N_+/N)/N)$ for the complementary log-log of the prevalence. (Contra Le, we do not recommend this as an estimator for the parameters, but it is a good choice of starting point for iterative methods.)

To examine the asymptotic properties of the pooled binomial distribution, we consider a series of experiments with increasing number of pools, $n$, where the empirical distribution of the pool sizes converges to a fixed distribution $p(s) = \lim_{n\to\infty}\hat{p}_n(s)$ (convergence in probability is sufficient here).[10] A sufficient condition for this convergence is for the sequence of pool sizes for the observations to be exchangeable, allowing application of the law-of-large-numbers. (To examine asymptotic behaviour we use our previous notation for the information function, but we will now put a subscript $n$ on the function to give explicit reference to the number of pools in the analysis.) Under this limit condition, we have limiting values for the asymptotic mean information matrices:

$$\bar{\mathbf{I}}(\theta, \lambda) \equiv \lim_{n \to \infty} \frac{\mathbf{I}_n(\theta, \lambda)}{n} \qquad \bar{\mathbf{I}}(\eta, \lambda) \equiv \lim_{n \to \infty} \frac{\mathbf{I}_n(\eta, \lambda)}{n}.$$

---

[9] Here we use the shorthand notation $\phi_s(\eta, \lambda) = 1 - \exp(-s\exp(\eta))$ (a slight abuse of notation since it uses the function $\phi_s$ in a generic sense that can be parameterised in different ways).

The mean information matrices $\bar{\mathbf{I}}(\theta, \lambda)$ and $\bar{\mathbf{I}}(\eta, \lambda)$ have the same form as the information matrices, replacing sums over pool sizes $s_i$ with expectations over the random pool size $S \sim p$.

The pooled binomial distribution obeys the standard regularity conditions that give asymptotic normality of the MLE. Specifically, as $n \to \infty$ we have convergence in distribution:

$$\sqrt{n} \begin{bmatrix} \hat{\theta}_{\text{MLE}} - \theta \\ \hat{\lambda}_{\text{MLE}} - \lambda \end{bmatrix} \xrightarrow{\text{Dist}} \text{N}(\mathbf{0}, \bar{\mathbf{I}}(\theta, \lambda)^{-1}),$$

or for the alternative parameterisation we have:

$$\sqrt{n} \begin{bmatrix} \hat{\eta}_{\text{MLE}} - \eta \\ \hat{\lambda}_{\text{MLE}} - \lambda \end{bmatrix} \xrightarrow{\text{Dist}} \text{N}(\mathbf{0}, \bar{\mathbf{I}}(\eta, \lambda)^{-1}).$$

For large $n$ this gives the approximating distributions:

$$\begin{bmatrix} \hat{\theta}_{\text{MLE}} - \theta \\ \hat{\lambda}_{\text{MLE}} - \lambda \end{bmatrix} \overset{\text{Approx}}{\sim} \text{N}(\mathbf{0}, \mathbf{I}_n(\theta, \lambda)^{-1}) \qquad \begin{bmatrix} \hat{\eta}_{\text{MLE}} - \eta \\ \hat{\lambda}_{\text{MLE}} - \lambda \end{bmatrix} \overset{\text{Approx}}{\sim} \text{N}(\mathbf{0}, \mathbf{I}_n(\eta, \lambda)^{-1}).$$

These large-sample distribution approximations for the MLEs of the parameters can be used to form confidence intervals for the true parameter values or formally test hypotheses pertaining to the parameter values. There is a substantial body of statistical literature examining various confidence intervals for the prevalence $\theta$ in the case where there is no dilution or intensification in the distribution (i.e., where we fix $\lambda = 0$ in the model). In this context, there are confidence intervals for the prevalence in Bhattacharyya, Karadinos and DeFoliart (1979), Hepworth (1996) and Hepworth (2005). In particular, the latter paper compares a range of confidence interval procedures including intervals based on the parameter $\eta$ from the complementary-log-log transformation; these confidence intervals are shown to produce satisfactory coverage results. Other confidence intervals for differences in prevalence parameters from independent samples are examined in Biggerstaff (2008).

---

[10] A sufficient condition here is to assume that the sequence of pool sizes is exchangeable (IID) and conditionally independent of the test results given the positive pool probability then the law of large numbers applies. In this case the law of large numbers applies, which ensures convergence. More generally, non-convergence will tend to occur only for "pathological" sequences, which we will assume are precluded. Our analysis here will continue to treat the pool sizes as being constants fixed by the design, so they are trivially independent of the test results.

5. Extension to allow covariate information

The above equations pertaining to the log-likelihood of the pooled binomial distribution apply in cases where we have pooled binomial data with the same underlying prevalence and intensity parameters. That analysis is framed in terms of a single vector observation from a distribution that involves multiple underlying binomial observations. In this section we examine extensions to the analysis involving multiple independent observations from the distribution, including cases where there is accompanying covariate information for the pools.

Extension to multiple independent observations from the pooled binomial distribution leads to simple variations of the above results. If we have independent observations with common parameters $\theta$ and $\lambda$ then we have $\prod_k \text{PoolBin}(\mathbf{y}_k|\mathbf{n}_k, \mathbf{s}_k, \theta, \lambda) = \text{PoolBin}(\mathbf{y}|\mathbf{n}, \mathbf{s}, \theta, \lambda)$ where we simply concatenate $\mathbf{y} = (\mathbf{y}_k)$, $\mathbf{n} = (\mathbf{n}_k)$ and $\mathbf{s} = (\mathbf{s}_k)$ over the data vectors. (To obtain a minimal sufficient statistic we need to reduce the vectors by aggregating the pool counts and positive pool counts over duplicate elements in the resulting pool size vector. It is possible to reduce the vectors so that the elements of $\mathbf{s}$ are distinct, giving minimal sufficient statistic $\mathbf{y}$.) If we have independent observations with different parameters —treated as free parameters— then the log-likelihood function is now a sum over the index $i$ containing parts given by the equations above, and the partial derivatives are sums of the above forms. The only non-trivial extension occurs when we have independent observations with different prevalence and/or intensity parameters that are not treated as free parameters, but are instead determined by some set of underlying shared parameters and/or covariates.

One useful extension of this kind is to add covariates for the pools, where the covariates affect the prevalence parameter. In this case we have as observation $(\mathbf{y}, \mathbf{n}, \mathbf{s}, \mathbf{x})$ where we now have a covariate matrix $\mathbf{x}$ containing a vector of covariates $\mathbf{x}_i$ for each pool. The covariates are taken to determine the prevalence parameter through a parametric equation $\theta_i = f(\mathbf{x}_i, \boldsymbol{\beta})$. Following Farrington (1992) we use the complementary log-log link function to obtain the link equation $\text{CLL}(\theta_i) = \eta_i = \mathbf{x}_i^\text{T} \boldsymbol{\beta}$ so $\text{CLL}(\phi_{s_i}(\theta_i, \lambda)) = (1 + \lambda) \log(s_i) + \mathbf{x}_i^\text{T} \boldsymbol{\beta}$. This gives a generalised linear model (GLM) where the pool size now can be viewed as one of the covariates in the model. This leads to an extended form of the pooled binomial distribution which we define below.

**DEFINITION (Pooled binomial distribution with covariates):** This is a simple extension of the pooled binomial distribution that allows covariate information on pools. The distribution has the mass function:

$$\text{PoolBin}(\mathbf{y}|\mathbf{n}, \mathbf{s}, \mathbf{x}, \boldsymbol{\beta}, \lambda) = \prod_i \text{Bin}(y_i|n_i, 1 - \exp(-s_i^{1+\lambda} \exp(\mathbf{x}_i^T \boldsymbol{\beta})))$$

$$= \exp\left(-\sum_i (n_i - y_i) s_i^{1+\lambda} \exp(\mathbf{x}_i^T \boldsymbol{\beta})\right) \prod_i \binom{n_i}{y_i} (1 - \exp(-s_i^{1+\lambda} \exp(\mathbf{x}_i^T \boldsymbol{\beta})))^{y_i},$$

where $\mathbf{n}$ is the **pool-count vector**, $\mathbf{s}$ is the **pool-size vector**, $\mathbf{x}$ is the **covariate matrix**, $\boldsymbol{\beta} \in \mathbb{R}^m$ is the **coefficient vector** and $\lambda \geq -1$ is the **excess intensity**. The prevalence parameters for the pools are defined by the transformation $\theta_i = \text{ICLL}(\mathbf{x}_i^T \boldsymbol{\beta}) = 1 - \exp(-\exp(\mathbf{x}_i^T \boldsymbol{\beta}))$. This distribution is a generalisation of the pooled binomial distribution without covariates.[11]

**REMARK:** This distribution accommodates cases where the covariates differ across the pools, but it does not allow covariates to differ for units within the same pool. Any experimental designs with different covariates for units within the same pool means that units in a pool may have different probabilities of having the marker of interest, and this changes the form of the positive pool probabilities, which goes outside the scope of the pooled binomial distribution (for examples of regression models that incorporate attributed of individuals as covariates, see e.g., Zhang et al 2013, Joyner et al 2020).

The log-likelihood function for this generalised distribution is a relatively simple extension of the results in the previous section. Taking the covariates as fixed values, the log-likelihood function is:

$$\ell_{\mathbf{y}|\mathbf{x}}(\boldsymbol{\beta}, \lambda) = \text{const} - \sum_i (n_i - y_i) s_i^{1+\lambda} \exp(\mathbf{x}_i^T \boldsymbol{\beta}) + \sum_i y_i \log 1\text{mexp}(s_i^{1+\lambda} \exp(\mathbf{x}_i^T \boldsymbol{\beta})).$$

---

[11] We can recover the standard form of the pooled binomial distribution (without covariates) by taking $\mathbf{x}_i = 1$ for all pools and $\beta_0 = \text{CLL}(\theta)$ (the intercept term $\beta_0$ is then equivalent to the parameter $\eta$ in the analysis presented in the previous section). The prevalence parameter is recovered by the transform $\theta = 1 - \exp(-\exp(\beta_0))$ (i.e., the prevalence is the inverse complementary log-log of the intercept term in the model.

For brevity we will now write $\phi_i = 1 - \exp(-s_i^{1+\lambda} \exp(\mathbf{x}_i^T \boldsymbol{\beta}))$ as a shorthand to represent the positive pool probabilities for each pool in the analysis (noting that all these probabilities depend implicitly on the parameter vector $\boldsymbol{\beta}$). Differentiating with respect to $\boldsymbol{\beta}$ and applying the chain rule gives the score function and Hessian matrix:

$$\nabla_{\boldsymbol{\beta}} \ell_{\mathbf{y}|\mathbf{x}}(\boldsymbol{\beta}, \lambda) = -\sum_i \mathbf{x}_i \exp(\mathbf{x}_i^T \boldsymbol{\beta}) s_i^{1+\lambda} \left( n_i - \frac{y_i}{\phi_i} \right),$$

$$\nabla_{\boldsymbol{\beta}}^2 \ell_{\mathbf{y}|\mathbf{x}}(\boldsymbol{\beta}, \lambda) = -\sum_i \mathbf{x}_i \mathbf{x}_i^T \exp(\mathbf{x}_i^T \boldsymbol{\beta}) s_i^{1+\lambda} \left( n_i - \frac{y_i}{\phi_i} \left[ 1 - s_i^{1+\lambda} \exp(\mathbf{x}_i^T \boldsymbol{\beta}) \cdot \frac{1 - \phi_i}{\phi_i} \right] \right).$$

For any excess intensity value $\lambda$ the Hessian matrix for this function is negative semi-definite and so the MLE for $\boldsymbol{\beta}$ occurs at a critical point defined by the score equation $\nabla_{\boldsymbol{\beta}} \ell_{\mathbf{y}|\mathbf{x}}(\widehat{\boldsymbol{\beta}}, \lambda) = \mathbf{0}$ (which may not be unique if there is redundancy of information in the covariates). The corresponding score equation for the MLE $\hat{\lambda}$ is just as in the above equations. The MLE can be found using Newton iteration or via iteratively reweighted least squares. Once we have obtained MLEs $\widehat{\boldsymbol{\beta}}_{\text{MLE}}$ and $\hat{\lambda}_{\text{MLE}}$ we can obtain corresponding MLEs for the prevalence value in each pool:

$$\hat{\theta}_i = 1 - \exp(-\exp(\mathbf{x}_i^T \widehat{\boldsymbol{\beta}}_{\text{MLE}})).$$

The true prevalence values and their estimates both depend on the covariates for the pool, so in general, the distribution with covariates does not have a single prevalence value. Usually the inference for the prevalence will involve computation of confidence intervals for each pool in the sample (and possibly for new pools with known covariates and pool sizes). However, in some applications it may be natural to centre the covariates in the covariate matrix to give a "neutral" set of values, where the resulting estimated prevalence value can be considered as the estimated prevalence for an "average" sampling unit.[12]

Since the pooled binomial distribution is the distribution of an underlying set of independent binomial random variables, this distribution is a binomial GLM. It can be written as:

$$p(\mathbf{y}|\mathbf{n}, \mathbf{s}, \mathbf{x}, \boldsymbol{\beta}, \lambda) = \prod_i \text{Bin}(y_i|n_i, \phi_i) \qquad \text{CLL}(\phi_i) = (1 + \lambda) \log(s_i) + \mathbf{x}_i^T \boldsymbol{\beta}.$$

---

[12] This will often be the case in applications where all covariates are continuous variables on at least an interval scale —e.g., the covariates may be lengths, heights or weights of tested organisms. In such cases the mean values of the covariates represent average values across the sample. If the covariate matrix is centred then input values of zero for the centred covariates represent average values, and the resulting prevalence is for an "average" unit defined by having covariates that are all at their sample mean. Note that the non-linearity of the transformation from the covariates to the prevalence means that this prevalence of an "average unit" is not the same as the average of the prevalences taken across all the sample units.

We can see that the general form of the pooled binomial distribution (with or without covariates for the pools) arises from a standard binomial GLM using the complementary log-log function with the log-pool-sizes as a necessary covariate. The log-pool-sizes are covariates for the pools and the total intensity (i.e., $1 + \lambda$) is the "slope coefficient" for the effect of the log-pool-sizes. This equivalence allows us to bring in the general theory of GLMs to assist with model fitting, parametric inference, diagnostic testing, and other aspects of statistical analysis. (Later we will analyse an example of some pooled binomial data using some custom `R` functions to fit the generalised pooled binomial distribution; these functions are programmed as "wrappers" using general functions for GLMs in the underlying code.)

Extension of the pooled binomial distribution to accommodate covariates complicates the form of the Fisher information, but the exact form is not particularly illuminating, and is used only to obtain estimated standard errors for the parameter estimators. As with other aspects of the distribution, standard results for GLMs apply here, and so it is usual to rely on computational methods programmed for broad classes of GLMs. The fact that the pool size is separable from the other covariates under the model form means that it is not necessary to take account of the covariates in consideration of the "optimal" pool size.

One final remark is in order regarding the accurate estimation of the prevalence parameter and the model assumptions of pooled testing. In most applications, the object of interest is the prevalence parameter (or multiple prevalence values if there are covariates) and the intensity parameter is a nuisance parameter, which is included only if we want to allow scope for dilution or intensification in the testing mechanism. In terms of our GLM form, our inferential object of interest is the prevalence estimates for individual units as a function of their covariates (i.e., the estimates $\hat{\theta}_1, \ldots, \hat{\theta}_n$). The effect of the pool-size is usually a "nuisance" that is determined by the "slope term" measured by the estimated excess intensity. Consequently, the accuracy of our inference about the prevalence is heavily affected by whether or not we allow for the possibility of dilution or intensification in testing. If we exclude this possibility then we can fix $\lambda = 0$ in the model which fixes the slope term; this increases the accuracy of our estimate of the intercept term. Contrarily, if we allow for the possibility of dilution or intensification then we treat $\lambda$ as an unknown nuisance parameter to be estimated; this reduces the accuracy of our estimate of the intercept term. In practice, if there is insufficient variation in the pool sizes in the experiment then this difference can be quite large — indeed, if there is only one pool size in the model then the excess intensity is unidentifiable and so the other parameters in the model also become unidentifiable. Except for cases where there is a large coverage of many different pool sizes (where accurate estimation of the excess intensity is then possible), excluding the possibility of dilution or intensification may lead to a narrow interval for the prevalence, whereas including this possibility leads to a much wider interval. This is intuitively unsurprising —dilution and intensification of the testing mechanism can drastically change the effect of the pool size, which makes it much harder to estimate the prevalence accurately with pooled testing.

## 6. Diagnostic analysis of the effect of pool size

The assumptions of the pooled testing model give us a stipulated form for the function $\phi_s$, which gives the positive pool probabilities based on the underlying prevalence parameter. The assumed model form for pooled testing manifests in a positive pool probability "curve" that follows the equation $\text{CLL}(\phi_s) = (1 + \lambda) \log(s) + \text{covariate effect}$. If the model assumptions are wrong then this functional form may be incorrect, and it is desirable to test whether or not the stipulated form is falsified by the data. Since the pooled binomial distribution is obtained from an underlying binomial GLM, all the standard diagnostic tests for GLMs are applicable, but the form of the positive pool probabilities is of particular interest in this context.

One obvious benefit of our two-parameter pooled binomial distribution is that it already has a parameter that allows us to include dilution or intensification of the testing mechanism in the analysis. It is simple to test whether there is evidence of dilution or intensification via:

$$H_0: \lambda = 0 \qquad H_A: \lambda \neq 0.$$

Standard output for fitting the pooled binomial distribution using a GLM fit will include the coefficient estimates and estimated standard errors, so it is simple to conduct this hypothesis test using the asymptotic distribution of the MLE. In many cases, the analyst will have prior evidence that the testing mechanism does not suffer from dilution or intensification.[13] In some cases an analyst may wish to use a "hybrid" fitting method, where they first fit the model with variable excess intensity and then refit with a fixed intensity (with no dilution or intensification) if the hypothesis test indicates no significant evidence of this.

---

[13] Indeed, in practice, it would be highly unusual for a testing mechanism to have intensification when detecting a marker of interest. We include this possibility in our distributional form largely for completeness, since it does not require any additional model parameter once we have allowed for dilution.

The above hypothesis test takes care of the possibilities of dilution or intensification in the testing mechanism, but it will not detect general deviations from the assumed functional form of the positive pool probability. A natural way to test the assumed model form is to compare the fitted model with an alternative "saturated model" where the pool size enters into the link equation as a categorical variable (i.e., with no constraint on its effect) instead of through the linear term using the log-pool-size. The saturated model has the form:

$$p_{\text{SAT}}(\mathbf{y}|\mathbf{n}, \mathbf{s}, \mathbf{x}, \boldsymbol{\beta}, \boldsymbol{\mu}) = \prod_i \text{Bin}(y_i|n_i, \phi_i) \qquad \text{CLL}(\phi_i) = \sum_s \mu_s \mathbb{I}(s_i = s) + \mathbf{x}_i^T \boldsymbol{\beta},$$

where each parameter $\mu_s$ represents the effect of pool size $s$ and the overall parameter vector $\boldsymbol{\mu}$ is defined over all pool sizes in the data. Since the pool size enters the saturated model as a categorical variable there is no constraint in the "pattern" of the values representing the effects of the pool sizes. Consequently, it is possible to compare the estimated form of the positive pool probabilities under the saturated model with the constrained form of the "curve" of these values under the pooled binomial distribution. If the unconstrained form departs from the curve "too much" (in an appropriate statistical sense) then the data falsifies the assumed form of the positive pool probabilities in the model. This comparison can be conducted with a standard ANOVA comparison for nested models, with formal testing for model departure done using a chi-squared test on the likelihood ratio statistic. This gives us a useful "diagnostic test" for the assumed form of the positive pool probabilities in the pooled binomial distribution.

Diagnostic analysis of the assumed model form can also be done graphically by constructing a **pool probability plot**. An example is shown in Figure 4 below. The top part of the plot shows the distribution of the pool sizes in the data. The bottom part of the plot shows the estimated positive pool probabilities under the pooled binomial distribution (the red curve) and under the saturated model (the black dots and their error bars). By overlaying these two estimates on a single plot —with error bars showing confidence intervals for the saturated model— we obtain a clear graphical depiction of the degree to which the unconstrained estimates of the positive pool probabilities differ from the constrained estimates under the pooled binomial distribution. The figure shown was constructed from simulated data from a pooled binomial model with no intensification or dilution, and with no covariates.[14]

---

[14] Our simulated data was generated for $n = 400$ random pools with pool sizes $s_1, \ldots, s_n \sim$ IID Pois($\mu = 20$) and with true parameters $\theta = 0.0384$ and $\lambda = 1$ (i.e., no intensification or dilution in the testing mechanism). There were no covariates for the simulated data.

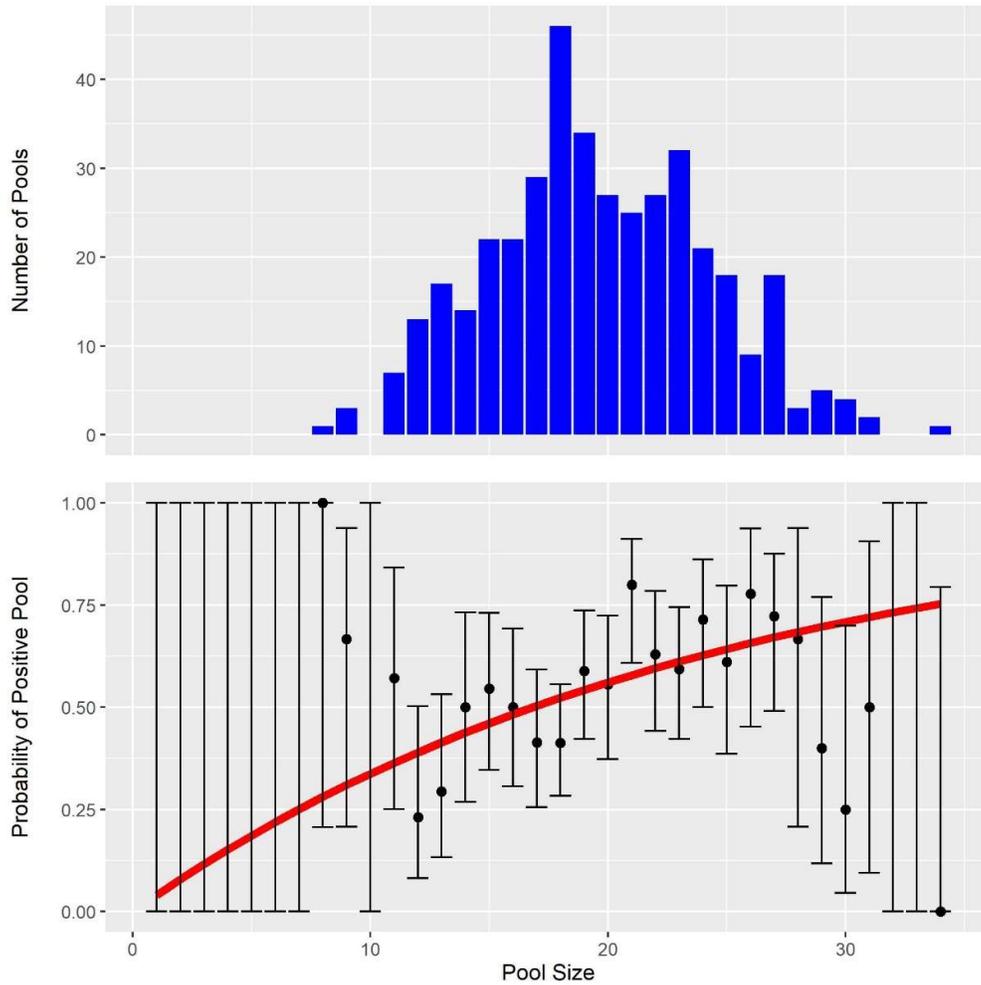

**FIGURE 4:** Pool probability plot from simulated pooled testing data

(Red line shows MLE; black points/bars show sample estimates for each pool size)

The pool probability plot shown here is for a simple simulated dataset with no covariates, but it can easily accommodate cases where there is covariate information on the pools. In either case the red line estimating the positive probability curve and the black estimates and error bars incorporate any covariate information through their effects on the parameter estimates in the model. One caveat to note in the pool probability plot is that the red curve formed using the MLE uses the data from all pools, so the estimate of this curve at a given pool size is not independent of the estimator for that pool size from the saturated model. In cases where a small number of pool sizes dominate the analysis there will be strong statistical dependence between the overall estimate in the red curve and the empirical estimates for the dominating pools, which makes it harder to diagnose departures from the model assumptions. Consequently, one useful variation for the plot is to adjust the estimated curve (red line) by estimating each point using the MLE of a dataset that *excludes* the data in that pool. This adjustment leads to an estimator that is independent of the empirical estimates, allowing easier diagnosis of departures from model assumptions.[15]

In the figure above we can see that fitting the pooled binomial distribution entails an attempt to fit the positive probability "curve" to the estimated points provided by the saturated model for each individual pool size. Looking at the estimation problem this way sheds some light on the degrees-of-freedom in the model, and the necessity for having variation in the pool sizes in the experimental design. For simplicity, let us consider the case where there is no covariate information. If there is only *one pool size* in the data then the two-parameter pooled binomial distribution gives an unidentifiable model — fitting the distribution here amounts to an attempt to fit a two-parameter curve through a single point, and so there is not enough information to simultaneously estimate the prevalence and excess intensity. If there are only *two pool sizes* in the data then the parameters are identifiable, but we cannot obtain standard error estimates for the estimators — fitting the distribution here amounts to fitting a two-parameter curve through two points, giving a perfect fit for the curve, subject only to the constraints on the parameter space.[16] It is only when we have at least three distinct pool sizes in the data that we can simultaneously estimate the prevalence and excess intensity parameters, and have at least one degree-of-freedom left over to give standard error estimates of the parameters.

As a practical matter, the simultaneous estimation of prevalence and excess intensity means that the estimate of the prevalence is subject to substantially larger standard error. In most analyses of pooled data, there is a large difference in the accuracy of estimates of the prevalence depending on whether we use the two-parameter model (with variable excess intensity) or a one-parameter model (with fixed excess intensity). This is intuitively unsurprising — allowing for variation in the excess intensity means that there can be substantial variation in the positive probability for large pool sizes without any change in the prevalence. This means that analysts have to pay close attention to their model assumptions.

---

[15] Note with this adjustment that the red line will no longer be as smooth curve; there will usually be "kinks" in the curve owing to the effect of the changing data.

[16] There is one case where we will not get a perfect fit. If the estimated points provided by the saturated model give an estimated positive probability that is higher for the lower pool size, then the estimated "curve" from the pooled binomial distribution will be a flat line with zero estimated total intensity (i.e., pool size having no effect on the positive pool probability). This occurs because the assumed form of the curve $\phi_s$ is non-decreasing, so it cannot perfectly fit points that are decreasing.

## 7. Application to mosquito data

In order to illustrate the application of our pooled testing distribution we will examine a pooled testing dataset contained in Table 1 of Walter, Hildreth and Beaty (1980). The data give results for pooled tests assaying yellow fever virus from a progeny population of *Aedes aegypti* (Santo Domingo strain) mosquitos whose parent population were infected with the virus. This dataset contains testing data on $N = 17,834$ sample units grouped into $K = 144$ pools, varying from pool sizes of $s = 5$ units up to $s = 287$ units.[17] The dataset includes two binary covariates, giving rise to four distinct groups of sample units: those with yellow-fever viral strains A and H; and those with larval development 6-10 days and 11-15 days.

The pooled testing distribution in this paper is implemented in the **pooltesting** package in **R** (O'Neill 2020). The pooled testing model can be fit to a dataset with or without covariates using the **pooltest** function in that package. In the code below we load the dataset and fit the data to the pooled testing model using the viral strain and larval development variables as covariates in the model. (Commands are shown in blue and output is shown in black; some parts of the output is replaced with ... for brevity.) The formula syntax for the function uses the positive pool counts as the response variable and the covariates as explanatory variables; the variable names for the pool counts and pool sizes are entered as separate variables in the function.

```
#Load the data and look at the first few rows
DATA <- readRDS("Walter et al Mosquito Data.rds")

#Model the data with variable intensity
TEST <- pooltest(positive ~ Virus + Development,
                 poolcount = pools, poolsize = poolsize,
                 data = DATA)
```

The output created by this function is a standard model object in **R**, and we can use all the standard commands for outputs of the model.[18] Once the model has been fit we can extract the coefficient estimates table (which gives estimates of the coefficients and their standard errors with tests of significance) and the ANOVA table. We can also extract an ANOVA comparison of the model with the unconstrained model where pool size is fit as a factor variable.

---

[17] This dataset is saved in the supplementary materials as the file **Walter Mosquito Data.rds**. The dataset contains the positive pool counts in the variable **positive**, the pool counts in the variable **pools** and the pool sizes in the variable **poolsize**.

[18] The output is an object with classes **'pooltest'**, **'glm'** and **'lm'**. This object contains standard model information, plus some additional information for the pooled binomial distribution. The package contains some custom functions for objects of type **'pooltest'** (e.g., to extract diagnostic information for the model).

```
summary(TEST)

...

Coefficients:
                    Estimate Std. Error  t value  Pr(>|t|)
(Intercept)          -5.8882     3.1954   -1.843   0.0704 .
ExcessIntensity      -0.3230     0.6858   -0.471   0.6394
VirusH                0.9189     0.5340    1.721   0.0905 .
Development11-15 days 1.4969     0.4360    3.433   0.0011 **
---
Signif. codes:  0 '***' 0.001 '**' 0.01 '*' 0.05 '.' 0.1 ' ' 1

...

anova(TEST)

Analysis of Deviance Table

...

                 Df   Deviance   Resid. Df   Resid. Dev
NULL                                   62       89.988
ExcessIntensity   1     0.1748         61       89.813
Virus             1     1.7588         60       88.054
Development       1    15.6459         59       72.408

diagnostic(TEST)

Analysis of Deviance Table (Actual Model vs Unconstrained Model)

...

                    Resid. Df  Resid. Dev  Df  Deviance  p-value
Actual Model              59      72.408
Unconstrained Model        9       3.186   50    69.222  0.037171
```

The above output shows the coefficient estimates and analysis of variance for the model, plus the analysis of variance comparison to the unconstrained model. The coefficient estimates allow us to estimate the prevalence for any of the four groups defined by the covariates, and also allows us to test whether there is intensification or dilution in the testing mechanism. From the diagnostic output comparing the actual model with the unconstrained model we have a p-value 0.0372 which means that there is some evidence of a departure from the stipulated form of the positive pool probabilities (with respect to the pool size variable). From the coefficient output for the **ExcessIntensity** term we have estimated excess intensity $\hat{\lambda} = -0.3230$ and the coefficient test has p-value 0.6394, so that —*within the assumed model form*—there is no evidence of dilution or intensification (i.e., of a non-zero value for the excess intensity).

Based on this test output we might decide to refit the model with the excess intensity parameter fixed at $\lambda = 0$ (i.e., no allowance for dilution or intensification in the model). The output of this latter model is shown below.

```
#Model the data with fixed intensity
TEST2 <- pooltest(positive ~ Virus + Development,
                  poolcount = pools, poolsize = poolsize,
                  data = DATA, fixed.intensity = TRUE)

summary(TEST2)

...

Coefficients:
                     Estimate Std. Error  t value  Pr(>|t|)
(Intercept)           -7.3752     0.5075  -14.532   < 2e-16 ***
VirusH                 0.8064     0.4822    1.672   0.09964 .
Development11-15 days  1.4651     0.4311    3.398   0.00121 **
---
Signif. codes:  0 '***' 0.001 '**' 0.01 '*' 0.05 '.' 0.1 ' ' 1

...

anova(TEST2)

Analysis of Deviance Table

...

            Df  Deviance  Resid. Df  Resid. Dev
NULL                            62      89.988
Virus        1    1.9293         61      88.058
Development  1   15.3606         60      72.698

diagnostic(TEST2)

Analysis of Deviance Table (Actual Model vs Unconstrained Model)

...

                     Resid. Df  Resid. Dev  Df  Deviance   p-value
Actual Model                60      72.698
Unconstrained Model          9       3.186  51    69.512  0.043347
```

Again, the coefficient estimates allow us to estimate the prevalence for any of the four groups defined by the covariates. Since the model now fixes the excess intensity parameter (with no allowance for dilution or intensification) the accuracy of our inference about the prevalence has increased, as indicated by the lower estimated standard errors for the remaining model terms. From the diagnostic output comparing the actual model with the unconstrained model we have a p-value 0.0433 which means that there is still some evidence of a departure from the stipulated form of the positive pool probabilities (with respect to the pool size variable).

The above analysis of pooled testing data was conducted using the `pooltesting` package (O'Neill 2020), which was created to implement simple user-friendly analysis of the present pooled testing distribution. Other `R` packages for pooled testing are available which go beyond the present model form and methods (e.g., allowing variation in covariates for sample items in the same pools, allowing Bayesian estimation, etc.). Other notable packages for the analysis of pooled data include `binGroup` (Zhang et al 2018), `binGroup2` (Hitt et al 2020) and `PoolTestR` (McLure et al 2020). All these packages are useful for pooled testing — the recommended package for particular projects will depend largely on the desired model form and analysis method.

## 8. Concluding remarks

In this paper we have examined a general form for the pooled binomial distribution, arising from pooled testing of binary sampling units. For the case without covariates, we gave a two-parameter distribution with a prevalence parameter and an excess intensity parameter. For the case with covariates, we gave a more general form where the prevalence parameter is replaced with a vector of coefficients that operate on the covariates to determine the prevalence. In either case, the distributional form is a special case of the binomial GLM using the complementary-log-log link function and so it is amenable to general methods for GLM.

In our examination of this distribution we have seen that the information properties confirm our intuitive idea that pooled testing entails a loss of information in individual sampling units, with a larger loss of information when these units are placed in pools of larger size. We also saw that pooled testing may be beneficial in cases where the cost of testing is substantially higher than the cost of sampling, leading to a situation where the "information-per-unit-cost" is higher for larger pools. Analysis of the information function for the distribution allows us to obtain some simple heuristic results on the "optimal" pool size, but noting that we should have variation in pool sizes to allow adequate diagnostic testing of the model.

We have examined model-fitting and diagnostic testing of this distribution, both in theoretical terms, and using an applied example using the `pooltesting` package in `R`. We saw that the model form makes it easy to test for dilution or intensification in the testing mechanism, and it is also relatively simple to conduct diagnostic tests for departures from the model assumptions about the "curve" of the positive pool probabilities taken with respect to the pool size. The pooled binomial distributions examined here are useful distributions for pooled testing, and they allow simple and effective analysis of prevalence rates in pooled testing.

# Appendix

In this appendix we give derivations of results shown in the main body of the paper. We begin by showing the derivatives of the log-likelihood function from the pooled binomial distribution. The partial derivatives of $\phi_s(\theta, \lambda)$ up to second order are:

$$\frac{\partial \phi_s}{\partial \theta}(\theta, \lambda) = s^{1+\lambda}(1-\theta)^{s^{1+\lambda}-1},$$

$$\frac{\partial \phi_s}{\partial \lambda}(\theta, \lambda) = -s^{1+\lambda} \log(s) \log(1-\theta)(1-\theta)^{s^{1+\lambda}},$$

$$\frac{\partial^2 \phi_s}{\partial \theta^2}(\theta, \lambda) = -s^{1+\lambda}(s^{1+\lambda}-1)(1-\theta)^{s^{1+\lambda}-2},$$

$$\frac{\partial^2 \phi_s}{\partial \lambda^2}(\theta, \lambda) = -s^{1+\lambda} \log^2(s) \log(1-\theta)(1-\theta)^{s^{1+\lambda}}(1 + s^{1+\lambda} \log(1-\theta)),$$

$$\frac{\partial^2 \phi_s}{\partial \theta \partial \lambda}(\theta, \lambda) = s^{1+\lambda} \log(s)(1-\theta)^{s^{1+\lambda}-1}(1 + s^{1+\lambda} \log(1-\theta)).$$

The log-likelihood function for the distribution can be written as:

$$\ell_{\mathbf{y}}(\theta, \lambda) = \text{const} + \log(1-\theta) \sum_i (n_i - y_i) s_i^{1+\lambda} + \sum_i y_i \log \phi_{s_i}(\theta, \lambda).$$

The first partial derivatives of the log-likelihood function are:

$$\frac{\partial \ell_{\mathbf{x}}}{\partial \theta}(\theta, \lambda) = -\frac{1}{1-\theta} \sum_i (n_i - y_i) s_i^{1+\lambda} + \sum_i y_i \frac{s_i^{1+\lambda}(1-\theta)^{s_i^{1+\lambda}-1}}{\phi_{s_i}(\theta, \lambda)}$$

$$= -\frac{1}{1-\theta} \sum_i s_i^{1+\lambda} \left[(n_i - y_i) - y_i \frac{(1-\theta)^{s_i^{1+\lambda}}}{\phi_{s_i}(\theta, \lambda)}\right]$$

$$= -\frac{1}{1-\theta} \sum_i s_i^{1+\lambda} \left(n_i - \frac{y_i}{\phi_{s_i}(\theta, \lambda)}\right),$$

$$\frac{\partial \ell_{\mathbf{x}}}{\partial \lambda}(\theta, \lambda) = \log(1-\theta) \left[\sum_i (n_i - y_i) s_i^{1+\lambda} \log(s_i) - \sum_i y_i \frac{s_i^{1+\lambda} \log(s_i)(1-\theta)^{s_i^{1+\lambda}}}{\phi_{s_i}(\theta, \lambda)}\right]$$

$$= \log(1-\theta) \sum_i s_i^{1+\lambda} \log(s_i) \left[(n_i - y_i) - y_i \frac{(1-\theta)^{s_i^{1+\lambda}}}{\phi_{s_i}(\theta, \lambda)}\right]$$

$$= \log(1-\theta) \sum_i s_i^{1+\lambda} \log(s_i) \left(n_i - \frac{y_i}{\phi_{s_i}(\theta, \lambda)}\right).$$

The second partial derivatives of the log-likelihood function are:

$$\frac{\partial^2 \ell_{\mathbf{x}}}{\partial \theta^2}(\theta, \lambda) = -\frac{\partial}{\partial \theta} \frac{1}{1-\theta} \sum_i s_i^{1+\lambda} \left( n_i - \frac{y_i}{\phi_{s_i}(\theta, \lambda)} \right)$$

$$= -\frac{1}{(1-\theta)^2} \sum_i s_i^{1+\lambda} \left( n_i - \frac{y_i}{\phi_{s_i}(\theta, \lambda)} \right) - \frac{1}{1-\theta} \sum_i s_i^{1+\lambda} \left( \frac{y_i s_i^{1+\lambda}(1-\theta)^{s_i^{1+\lambda}-1}}{\phi_{s_i}(\theta, \lambda)^2} \right)$$

$$= -\frac{1}{(1-\theta)^2} \sum_i s_i^{1+\lambda} \left( n_i - \frac{y_i}{\phi_{s_i}(\theta, \lambda)} + \frac{y_i s_i^{1+\lambda}(1-\theta)^{s_i^{1+\lambda}}}{\phi_{s_i}(\theta, \lambda)^2} \right)$$

$$= -\frac{1}{(1-\theta)^2} \sum_i s_i^{1+\lambda} \left( n_i - \frac{y_i}{\phi_{s_i}(\theta, \lambda)} \left[ 1 - \frac{s_i^{1+\lambda}(1-\theta)^{s_i^{1+\lambda}}}{\phi_{s_i}(\theta, \lambda)} \right] \right)$$

$$= -\frac{1}{(1-\theta)^2} \sum_i s_i^{1+\lambda} \left( n_i - \frac{y_i}{\phi_{s_i}(\theta, \lambda)} \left[ 1 - s_i^{1+\lambda} \cdot \frac{1 - \phi_{s_i}(\theta, \lambda)}{\phi_{s_i}(\theta, \lambda)} \right] \right),$$

$$\frac{\partial^2 \ell_{\mathbf{x}}}{\partial \lambda^2}(\theta, \lambda) = \log(1-\theta) \sum_i \left[ \begin{array}{c} s_i^{1+\lambda} \log^2(s_i) \left( n_i - \frac{y_i}{\phi_{s_i}(\theta, \lambda)} \right) \\ -s_i^{1+\lambda} \log(s_i) \left( \frac{y_i s_i^{1+\lambda} \log(s_i) \log(1-\theta) (1-\theta)^{s_i^{1+\lambda}}}{\phi_{s_i}(\theta, \lambda)^2} \right) \end{array} \right]$$

$$= \log(1-\theta) \sum_i s_i^{1+\lambda} \log^2(s_i) \left[ n_i - \frac{y_i}{\phi_{s_i}(\theta, \lambda)} - \frac{y_i s_i^{1+\lambda} \log(1-\theta)(1-\theta)^{s_i^{1+\lambda}}}{\phi_{s_i}(\theta, \lambda)^2} \right]$$

$$= \log(1-\theta) \sum_i s_i^{1+\lambda} \log^2(s_i) \left( n_i - \frac{y_i}{\phi_{s_i}(\theta, \lambda)} \left[ 1 + \frac{s_i^{1+\lambda} \log(1-\theta)(1-\theta)^{s_i^{1+\lambda}}}{\phi_{s_i}(\theta, \lambda)} \right] \right)$$

$$= \log(1-\theta) \sum_i s_i^{1+\lambda} \log^2(s_i) \left( n_i - \frac{y_i}{\phi_{s_i}(\theta, \lambda)} \left[ 1 + s_i^{1+\lambda} \log(1-\theta) \cdot \frac{1 - \phi_{s_i}(\theta, \lambda)}{\phi_{s_i}(\theta, \lambda)} \right] \right),$$

$$\frac{\partial^2 \ell_{\mathbf{x}}}{\partial \theta \partial \lambda}(\theta, \lambda) = -\frac{1}{1-\theta} \sum_i s_i^{1+\lambda} \log(s_i) \left[ \left( n_i - \frac{y_i}{\phi_{s_i}(\theta, \lambda)} \right) - \frac{y_i s_i^{1+\lambda} \log(1-\theta)(1-\theta)^{s_i^{\lambda}}}{\phi_{s_i}(\theta, \lambda)^2} \right]$$

$$= -\frac{1}{1-\theta} \sum_i s_i^{1+\lambda} \log(s_i) \left( n_i - \frac{y_i}{\phi_{s_i}(\theta, \lambda)} \left[ 1 + \frac{s_i^{1+\lambda} \log(1-\theta)(1-\theta)^{s_i^{1+\lambda}}}{\phi_{s_i}(\theta, \lambda)} \right] \right)$$

$$= -\frac{1}{1-\theta} \sum_i s_i^{1+\lambda} \log(s_i) \left( n_i - \frac{y_i}{\phi_{s_i}(\theta, \lambda)} \left[ 1 + s_i^{1+\lambda} \log(1-\theta) \cdot \frac{1 - \phi_{s_i}(\theta, \lambda)}{\phi_{s_i}(\theta, \lambda)} \right] \right).$$

We now derive the Fisher information matrix for the distribution. As a preliminary step, we can confirm that the expected value of the score function (taken as random variable) is zero. Since $\mathbb{E}(y_i) = n_i \phi_{s_i}(\theta, \lambda)$ we have:

$$\mathbb{E}\left(n_i - \frac{y_i}{\phi_{s_i}(\theta,\lambda)}\right) = n_i - \frac{n_i \phi_{s_i}(\theta,\lambda)}{\phi_{s_i}(\theta,\lambda)} n_i - n_i = 0.$$

Using the above partial derivatives, it follows immediately that:

$$\mathbb{E}\left(\frac{\partial \ell_\mathbf{x}}{\partial \theta}(\theta,\lambda)\right) = \mathbb{E}\left(\frac{\partial \ell_\mathbf{x}}{\partial \lambda}(\theta,\lambda)\right) = 0.$$

The elements of the **Fisher information** matrix are:[19]

$$I_{\theta\theta}(\theta,\lambda) \equiv -\mathbb{E}\left(\frac{\partial^2 \ell_\mathbf{x}}{\partial \theta^2}(\theta,\lambda)\right)$$

$$= \frac{1}{(1-\theta)^2} \sum_i n_i s_i^{2(1+\lambda)} \cdot \frac{1-\phi_{s_i}(\theta,\lambda)}{\phi_{s_i}(\theta,\lambda)},$$

$$I_{\theta\lambda}(\theta,\lambda) \equiv -\mathbb{E}\left(\frac{\partial^2 \ell_\mathbf{x}}{\partial \theta \partial \lambda}(\theta,\lambda)\right)$$

$$= \log^2(1-\theta) \sum_i n_i s_i^{2(1+\lambda)} \log^2(s_i) \cdot \frac{1-\phi_{s_i}(\theta,\lambda)}{\phi_{s_i}(\theta,\lambda)},$$

$$I_{\lambda\lambda}(\theta,\lambda) \equiv -\mathbb{E}\left(\frac{\partial^2 \ell_\mathbf{x}}{\partial \lambda^2}(\theta,\lambda)\right)$$

$$= -\frac{\log(1-\theta)}{1-\theta} \sum_i n_i s_i^{2(1+\lambda)} \log(s_i) \cdot \frac{1-\phi_{s_i}(\theta,\lambda)}{\phi_{s_i}(\theta,\lambda)}.$$

This confirms the form of the Fisher information matrix shown in the main body of the paper. The corresponding unit information for $\theta$ (for a single sample unit in a pool of size $s$) is:

$$I_s(\theta,\lambda) = \frac{s^{2\lambda+1}(1-\theta)^{s^{1+\lambda}-2}}{1-(1-\theta)^{s^{1+\lambda}}}.$$

(It is then simple to establish that $\mathbf{I}_{\theta\theta}(\theta,\lambda) = \sum_k s_k \cdot I_{s_k}(\theta,\lambda)$ so that the Fisher information for $\theta$ is the sum of the unit information for $\theta$ taken over the entire sample.) In Table 1 below we show the lower prevalence cut-off values for "optimal" pool sizes in the case where we have no sampling cost and unit testing cost (where optimality is here determined by maximising the information-per-unit cost). This table shows the lower prevalence cut-offs for optimal pool sizes up to $\hat{s} = 40$.

---

[19] In the special case where $\mathbf{n} = n$ and $\mathbf{s} = 1$ we have $\text{PoolBin}(\mathbf{n},\mathbf{s},\theta,\lambda) = \text{Bin}(n,\theta)$ and the Fisher information for $\theta$ reduces down to the standard form for the binomial distribution, which is $I_{\theta\theta}(\theta) = n/\theta(1-\theta)$.

| Optimising pool size | Prevalence cut-off level | Optimising pool size | Prevalence cut-off level |
|---|---|---|---|
| 1 | 0.666682 | 21 | 0.071436 |
| 2 | 0.475310 | 22 | 0.068371 |
| 3 | 0.367461 | 23 | 0.065558 |
| 4 | 0.299097 | 24 | 0.063006 |
| 5 | 0.252064 | 25 | 0.060612 |
| 6 | 0.217754 | 26 | 0.058394 |
| 7 | 0.191641 | 27 | 0.056333 |
| 8 | 0.171107 | 28 | 0.054412 |
| 9 | 0.154539 | 29 | 0.052618 |
| 10 | 0.140893 | 30 | 0.050938 |
| 11 | 0.129457 | 31 | 0.049363 |
| 12 | 0.119737 | 32 | 0.047882 |
| 13 | 0.111373 | 33 | 0.046487 |
| 14 | 0.104101 | 34 | 0.045172 |
| 15 | 0.097719 | 35 | 0.043928 |
| 16 | 0.092074 | 36 | 0.042752 |
| 17 | 0.087045 | 37 | 0.041636 |
| 18 | 0.082537 | 38 | 0.040578 |
| 19 | 0.078472 | 39 | 0.039572 |
| 20 | 0.074789 | 40 | 0.038615 |

**TABLE 1:** Pool size maximising information-per-unit-cost

(Prevalence values above the listed cut-off but below the above cut-off)

Finally, we confirm the monotonicity claim in the main body of the paper, which asserts that the unit information is strictly decreasing in $s$ for all $\theta > 0$ in the case where $\lambda = 0$ (i.e., no dilution or intensification). With a bit of algebra we can show that:

$$I_{s+1}(\theta, 0) - I_s(\theta, 0) = \frac{(1-\theta)^{s-2}}{\phi_{s+1}(\theta)\phi_s(\theta)} \left[ (1 - (s+1)\theta) - (1-\theta)^{s+1} \right].$$

We need to show that this quantity is negative for all $\theta > 0$ and $s \in \mathbb{N}$. To do this, we will use the well-known exponential inequality $e^{-x-x^2} \leq 1 - x \leq e^{-x}$. Applying this inequality gives:

$$1 - (s+1)\theta \leq e^{-(s+1)\theta} < e^{-(s+1)\theta} e^{-(s+1)\theta^2} \leq (1-\theta)^{s+1}.$$

Applying this latter inequality to the above expression shows that $I_{s+1}(\theta, 0) - I_s(\theta, 0) < 0$ for all $\theta > 0$ and $s \in \mathbb{N}$, which shows that $I_s(\theta, 0)$ is strictly decreasing in $s$. This establishes that the information from an individual sample participant is lower when the participant is in a pool with a larger pool size.

For brevity we omit the corresponding working for the reparameterised log-likelihood function $\ell_{\mathbf{y}}(\eta, \lambda)$. This working as analogous to the above, or it can be derived alternatively using the chain rule from the above results.